\documentclass[12pt,bibliography,preprint]{iopart}
\usepackage{graphicx}
\usepackage{epstopdf}

\begin{document}
\title{Optomechanics based on angular momentum exchange between light and matter}

\author{H. Shi}
\address{Department of Physics, G07 Physical Sciences Building, Cornell University, Ithaca,
NY 14853, USA}
\ead{hxs673@cornell.edu}
\author{M. Bhattacharya}
\address{School of Physics and Astronomy, Rochester Institute of Technology, 84 Lomb Memorial Drive,
Rochester, NY 14623, USA}
\ead{mxbsps@rit.edu}

\date{\today}

\begin{abstract}
The subject of optomechanics involves interactions between optical and mechanical degrees of freedom, and
is currently of great interest as an enabler of fundamental investigations in quantum mechanics, as well
as a platform for ultrasensitive measurement devices. The majority of optomechanical configurations rely
on the exchange of linear momentum between light and matter. We will begin this tutorial with a
brief description of such systems. Subsequently, we will introduce optomechanical systems based on
\emph{angular} momentum exchange. In this context, optical fields carrying polarization and
orbital angular momentum will be considered, while for the mechanics, torsional and free rotational motion
will be of relevance. Our overall aims will be to supply basic analyses of some of the existing theoretical
proposals, to provide functional descriptions of some of the experiments conducted thus far, and to consider
some directions for future research. We hope this tutorial will be useful to both theorists and experimentalists
interested in the subject.
\end{abstract}

\maketitle
\section{Introduction}

Optomechanical systems, which are based on the interaction between optical and mechanical degrees of
freedom, are currently under intense research focus
\cite{KarraiReview2006,VahalaReview2008,MarquardtReview2008,GirvinReview2009,ClerkReview2010,AspelmeyerReview2010,
McClellandReview2011,RegalReview2011,AspelmeyerReview2012,MeystreReview2013,AspelmeyerRMP2014,MetcalfeReview2014,
AranyaReview2014}. Much of the interest in these systems is due to their ability to access quantum mechanics at more
macroscopic scales than ever before. For example, a major accomplishment of the field
has been the cooling of a microfabricated harmonic oscillator, consisting of roughly $10^{11}$ atoms, to
its quantum mechanical ground state \cite{ChanNature2011}. Optomechanical systems therefore provide an
avenue for investigating fundamental quantum mechanical issues such as wave function collapse
\cite{NimmrichterPRL2014,DiosiPRL2015}, superposition \cite{MarshallPRL2003,RomeroIsartPRL2011}, decoherence
\cite{PepperPRL2012}, entanglement \cite{LeeScience2011,PalomakiScience2013}, backaction \cite{PurdyScience2013}
and Planck-scale physics \cite{PikovskiNaturePhysics2012}. Furthermore, these optomechanical systems hold
great potential as platforms for ultrasensitive measurement technologies. Indeed, optomechanical principles
underlie cutting-edge displacement \cite{HertzbergNaturePhysics2010} and force
\cite{SchrepplerScience2014,MoorePRL2014} sensors, phonon detectors \cite{ClerkPRL2010,CohenNature2015},
gravitational wave interferometers \cite{McClellandReview2011}, thermometers \cite{MillenNatureNano2014},
accelerometers \cite{KrauseNaturePhotonics2012}, and magnetometers \cite{ForstnerPRL2012}, to cite a few examples.
While a variety of physical configurations are used in the investigations mentioned above, in almost all of them,
the optical and mechanical degrees of freedom interact via the exchange of \textit{linear} momentum.

However, light can also exchange \textit{angular} momentum with matter
\cite{OAMBook,KolobovBook,BekshaevBook,TorresBook,AndrewsBook,Andrews2Book,Allen1999,Loudon2003,Padgett2004,Arnold2008,Yao2011,Bliokh2011}.
Since the pioneering experiment of Beth, it has been known that light can carry spin angular momentum (i.e.
polarization), which in turn can exert torque on bulk matter \cite{BethPR1936,MoothooAJP2001}. Subsequently, a seminal
paper by Allen et al. in 1992 pointed out that photons could also carry orbital angular momentum (OAM), encoded as
spatial structure in the transverse profile of the corresponding optical beam \cite{AllenPRA1992,GalvezNJP2011}.
Unlike spin, which can only take values $s=\pm \hbar$, photonic OAM can take any value $l=0,\pm \hbar, \pm 2\hbar,\ldots.$
Of importance to this tutorial is the fact that OAM-carrying beams can exert torque
on matter as well, with experiments demonstrating optical rotation effects on nanoparticles
\cite{HePRL1995,SimpsonOL1997,YoshiNatComm2013},
liquid crystals \cite{PiccirilloPRL2001}, and Bose-Einstein condensates \cite{AndersenPRL2006,RyuPRL2007,WrightPRA2008},
for example. Also of relevance is the fact that, due to their intrinsic spatial inhomogeneities, beams with OAM
can sense the circulation of material particles, via the rotational Doppler shift \cite{BirulaPRL1997,BarreiroPRL2006}.
This has been verified in velocimetry experiments \cite{YoshiNatComm2013,LaveryScience2013,PhillipsPRA2014,
LaveryOptica2014} (which have also been implemented with circularly polarized beams \cite{KorechNatPhot2013}).
Further, it is relevant to mention that the practically infinite Hilbert space corresponding to the virtually
unlimited OAM available per photon has proved to be an important resource for quantum information processing
\cite{Mair2001,Terriza2002,Collins2002,Cerf2002,TerrizaNatPhys2007,Gbur2008,Lanyon2009,Giovannini2011,Zeilinger2012,
Escartin2012,MirhosseiniNJP2015}. Lastly, OAM carried by an optical beam is associated with the
presence of singularities, i.e. vortices, in the beam profile. These structures have their own interesting
dynamics and place OAM-carrying beams into the larger arena of `singular optics'
\cite{BerryPR1974,MansuripurOPN1999,Swartzlander2006,Dennis2009}. There are many other applications of
OAM-carrying beams; we have mainly mentioned those which are relevant to our discussion.

This tutorial aims to introduce optomechanical systems which are based on angular momentum exchange between
light and matter. For the optical fields, this will imply the consideration of photons carrying either
polarization or OAM. For the mechanics, torsional and free rotational motion will be of importance. To our
knowledge, existing introductions to, and reviews of, the micromanipulation of matter with angular
momentum-carrying beams largely deal with liquid media \cite{Padgett1997,DholakiaPW2002,GrierNature2003,Moffitt2008}.
In contrast, the present tutorial will consider instead only systems under vacuum. With few exceptions, we
will consider beams carrying well-defined optical angular momentum per photon, and will neglect the effects of
linear photonic momentum on mechanical torsional or free rotational motion
\cite{Tittonen1999,Mueller2009,Wang2009,Kim2013}.

It is worth stating explicitly the interest behind considering the two types of mechanical motion mentioned above.
Torsional oscillators are often the instruments of choice for precision measurement experiments. For example,
they were used in historic measurements of the static electric force by Coulomb \cite{Coulomb1785}, of the
gravitational force by Cavendish \cite{Cavendish1798}, and of optical polarization by Beth \cite{BethPR1936}.
Furthermore, a key difference between linear and torsional oscillations is that the linear vibrational motion of
an object is affected by its total mass, while the torsional oscillations are affected not only by the total mass but
also by how that mass is distributed, i.e. by the moment of inertia. For example, the mass of a sphere of radius
$r$ varies as $r^{3}$, but the moment of inertia goes as $r^{5}.$ Thus, smaller objects are easier to move linearly,
but even easier to move torsionally. These distinctions make it worthwhile to study torsional oscillations on their
own.

On the other hand, free mechanical rotation presents some entirely
novel features, compared to linear or torsional oscillatory motion, especially in the quantum regime. For example,
it is well known that since the harmonic oscillator is a linear system (i.e. its equations of motion are linear
in the dynamical variables, or equivalently, the energy eigenvalues are evenly spaced), quantum effects are quite
difficult to observe. In other words, the quantum oscillator behaves, for the most part, like a classical
oscillator. To observe quantum effects, one usually needs to introduce some sort of nonlinear interaction, for example,
by coupling the oscillator to the mode of an electromagnetic field, or to a two-level system \cite{AspelmeyerRMP2014}.
The free rotor, in contrast, is a nonlinear system (i.e. its equations of motion are nonlinear in the dynamical
variables, or equivalently, its energy spectrum is anharmonic, see below), which suggests that it might be easier
to observe quantum effects in its behavior without the use of any auxiliary systems. One such quantum effect,
which would be interesting to verify at much larger than atomic mass scales, is the quantization of angular
momentum, one of the central predictions of quantum mechanics \cite{SakuraiBook}.
%Further, the free rotor is unusual
%in possessing no ground state energy, unlike all the other textbook examples of quantum systems, such
%as the finite and infinite square wells, the delta function potential, and the harmonic oscillator. Thus a
%mechanical rotor cooled to its ground state would represent a degree of freedom from which all energy,
%including that of quantum fluctuations, has been removed.
A second intriguing feature of the rotor is that, unlike the harmonic oscillator (and other elementary quantum
systems such as the square well, the delta function well, or the Morse potential), it has no ground state
energy \cite{SakuraiBook}. This can be seen from the Hamiltonian of a particle of mass $m$ rotating in a circle
of radius $r$,
\begin{equation}
\label{eq:HamRot}
H_{r}=\frac{L_{z}^{2}}{2I},
\end{equation}
where $L_{z}$ is the angular momentum of the particle, and
\begin{equation}I=mr^{2},
\end{equation}
is its moment of inertia. The eigenvalues of $H_{r}$ are $l^{2}\hbar^{2}/2I, l =0,1,2,\ldots.$ Thus, the ground
state $(l=0)$ is perfectly rotationless. Finally, the prospect of observing counter-rotating
superpositions of mechanically rotating states \cite{ShiJPhysB2013} is quite tantalizing, and is the neutral
mechanical analog of counter-propagating persistent electronic currents in superconducting circuits
\cite{Mooij1999}. From a more practical point of view, freely rotating mechanical systems could possibly be
used for gyroscopy \cite{YoshiNatComm2013} and for storing photonic OAM \cite{ShiJPhysB2013}. Conversely,
optical beams carrying OAM could be used to probe mechanical rotation, in a quantum-limited manner
\cite{JhaPRA2011}.

The remainder of the tutorial is organized in the following way. We first introduce optomechanics based on
electromagnetic modes confined to optical resonators. Section~\ref{sec:VCO} describes cavity-based vibrational
optomechanics, and lays the ground for the formalism, basic physics and applications to follow later.
Section~\ref{sec:OAM} then introduces OAM-carrying beams. The material presented thus far is then utilized to
explain cavity-based torsional and rotational optomechanics in Sections~\ref{sec:TorsionalOpto} and~\ref{sec:RCO}
respectively. Accounting for the more recent interest in cavityless optomechanical systems, we
have included some material on this topic in Section~\ref{sec:CavLess}. The tutorial ends with a
Conclusion in Section~\ref{sec:Con} and Acknowledgments in Section~\ref{sec:Ack}.

The format for each section is to first introduce a basic physical configuration. This is done
typically, but not in all cases, using work published by our own group, simply because we understand our
own work best. Subsequently, reference is made to other works in the field. Although we have tried to be inclusive
in making such references, we apologize to all authors whose works are not adequately cited.

\section{Vibrational cavity optomechanics}
\label{sec:VCO}
In this section, we will provide a brief review of standard linear vibrational cavity optomechanics \cite{ShiAJP2013};
those who are familiar with the material may wish to proceed directly to Section~\ref{sec:OAM}. A number of
comprehensive reviews of the topic already exist
\cite{VahalaReview2008,GirvinReview2009,MeystreReview2013,AspelmeyerRMP2014},
so we will not be aiming to achieve detail or rigor in our presentation. Rather, our discussion will be
largely heuristic, focusing on the physics rather than the mathematical details, and aimed mainly at making the
ensuing discussion of rotational optomechanics accessible. Although we will be making reference to both classical
and quantum regimes for the relevant optomechanical systems, we will largely employ the formalism of second
quantization, which makes the notation compact and also reveals the microscopic physics of the system in terms of
its fundamental quanta. We begin below by reviewing this formalism for massive particles as well as electromagnetic
fields.

\subsection{Second quantization for a massive harmonic oscillator}
Consider a simple harmonic oscillator of mass $m$ oscillating at frequency $\omega_{m}$, in one dimension. The
corresponding classical mechanical Hamiltonian is given by \cite{ShiAJP2013,SakuraiBook}
\begin{equation}
\label{eq:HamSHO}
H=\frac{p^{2}}{2m}+\frac{1}{2}m\omega_{m}^{2}q^{2},
\end{equation}
where $q$ and $p$ are the oscillator position and momentum, respectively. If we now consider the oscillator
to be a quantum object, the two dynamical variables must obey the commutation relation
\begin{equation}
\label{eq:CommCan1}
[q,p]=i\hbar,
\end{equation}
where $\hbar$ is Planck's constant divided by $2\pi$. The dynamical evolution of any quantum mechanical
operator $O$ can be found using the Heisenberg equation \cite{SakuraiBook}
\begin{equation}
\label{eq:HE}
\dot{O}= \frac{i}{\hbar}\left[H,O\right]+\frac{\partial O}{\partial t},
\end{equation}
where $t$ is time, the dot denotes a total time derivative, and the last term is nonzero only if the operator
carries explicit time-dependence. Using this equation the Heisenberg equations for $q$ and $p$ follow
from Eq.~(\ref{eq:HamSHO})
\begin{equation}
\label{eq:SHOQP1}
\dot{q} = \frac{p}{m},
\end{equation}
\begin{equation}
\label{eq:SHOQP2}
\dot{p} = -m\omega_{m}^{2}q.
\end{equation}
In real systems, and certainly in optomechanical setups, it is important to account for damping and
fluctuations experienced by the oscillator as a result of its coupling to the environment. The procedure
for incorporating such effects in a way consistent with quantum mechanics is somewhat complicated and we
will not describe it in detail. Fortunately this procedure results in a straightforward prescription
called the Quantum Langevin approach \cite{AspelmeyerRMP2014,GenesPRA2008}, which preserves
Eq.~(\ref{eq:SHOQP1}) but modifies Eq.~(\ref{eq:SHOQP2}), as
\begin{eqnarray}
\label{eq:SHOQLE}
\dot{q} &= \frac{p}{m}, \\
\dot{p} &= -m\omega_{m}^{2}q-\gamma_{m}p+\xi,
\end{eqnarray}
where $\gamma_{m}$ is the rate of mechanical damping and $\xi$ is a Brownian stochastic force with zero
mean which obeys the correlation function \cite{GenesPRA2008}
\begin{equation}
\label{eq:harmcorr1}
\langle \xi(t)\xi(t')\rangle = \frac{\gamma_{m}}{\omega_{m}}\int \frac{d\omega}{2\pi}e^{-i\omega(t-t')}
\omega\left[\coth\left(\frac{\hbar\omega}{k_{B}T}\right)+1\right],
\end{equation}
where $k_{B}$ is Boltzmann's constant and $T$ is the ambient temperature. Typically we will deal with
high temperatures such that $k_{B}T \gg \hbar\omega_{m}$. In this limit, the correlation simplifies to
\begin{equation}
\label{eq:harmcorr2}
\langle \xi(t)\xi(t')\rangle = \frac{2\gamma_{m}k_{B}T}{\hbar\omega_{m}}\delta(t-t').
\end{equation}
Working with the quantum mechanical variables $q$ and $p$ is convenient since they allow us to take the
classical limit of the theory when required. However, sometimes it is more convenient to use instead a set
of variables which describe the system in terms of its basic quanta. Introducing the transformations
\cite{ShiAJP2013}
\begin{equation}
\label{eq:qpTOaad}
q=q_{0}\left(b^{\dagger}+b\right), \,\,\, p=ip_{0}\left(b^{\dagger}-b\right),
\end{equation}
where
\begin{equation}
q_{0}=\sqrt{\frac{\hbar}{2m\omega_{m}}}, \,\,\,p_{0}=\sqrt{\frac{m\hbar\omega_{m}}{2}},
\end{equation}
the Hamiltonian of Eq.~(\ref{eq:HamSHO}) can be rewritten as
\begin{equation}
\label{eq:HamSQSHO}
H=\hbar \omega_{m}b^{\dagger}b,
\end{equation}
where we have dropped a constant offset $\hbar \omega_{m}/2,$ which corresponds to the ground state energy
of the oscillator, and has no dynamical effect. In this formalism, the operator $b (b^{\dagger})$ physically
corresponds to the destruction (creation) of one quantum of the oscillator energy, i.e. one phonon. The
combination $b^{\dagger}b$ is called the number operator, as its value denotes the number of quanta carried
by the oscillator, with the energy per quantum being $\hbar \omega_{m}$ as per Eq.~(\ref{eq:HamSQSHO}). From
Eqs.~(\ref{eq:CommCan1}) and ~(\ref{eq:qpTOaad}), it follows that
\begin{equation}
[b,b^{\dagger}]=1.
\end{equation}

%Using this equation the equation for $a$ follows
%\begin{equation}
%\label{eq:aHei}
%\dot{b}=i\omega_{m}b,
%\end{equation}
%where there is no contribution from the partial time derivative as $b$ carries no explicit time dependence.
%As can be verified readily, Eq.~(\ref{eq:aHei}) does have such a solution,
%\begin{equation}
%\label{eq:bsol}
%b(t)=b(0)e^{i\omega_{m}t}.
%\end{equation}

\subsection{Cavity quantization of an electromagnetic field mode}
\label{subsec:CavQ}
\begin{figure}
\begin{center}
\includegraphics[width=0.4\textwidth]{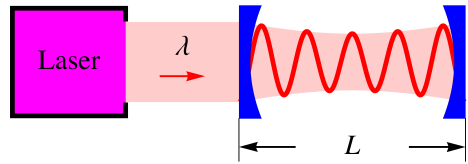}
\caption{(Color online) Single frequency cavity mode described in Section~\ref{subsec:CavQ}. A laser injects
light of wavelength $\lambda$ into a Fabry-Perot cavity made from two highly reflecting mirrors. The
cavity length is $L$.}
\label{fig:P1}
\end{center}
\end{figure}

We now consider the quantization of a mode of the electromagnetic field, following the approach of
Ref.~\cite{GerryBook}. By a mode we mean an electromagnetic wave of well-defined frequency $\omega_{0}$.
In order to isolate a single optical mode, it is convenient to use two mirrors facing each other, thus
realizing a Fabry-Perot cavity, as shown in Fig.\ref{fig:P1}. The classical electromagnetic Hamiltonian,
which combines the electric and magnetic contributions, is given by
\cite{GerryBook}
\begin{equation}
\label{eq:HamOP}
H_{0}=\frac{1}{2}\int dV \left[\epsilon_{0}E^{2}+\frac{B^{2}}{\mu_{0}}\right],
\end{equation}
where $E$ and $B$ are the electric and magnetic fields, respectively, and the integration is over the
cavity volume. As a useful idealization, the two mirrors, which are a length $L$ apart, are taken to be
perfectly reflecting. This assumption will be relaxed later. The quantized electric and magnetic fields
which satisfy the appropriate boundary conditions are
\begin{equation}
\label{eq:EL}
E_{x}(z,t)=E_{0}\left(a^{\dagger}+a\right)\sin kz,
\end{equation}
\begin{equation}
\label{eq:MA}
B_{y}(z,t)=iB_{0}\left(a^{\dagger}-a\right)\cos kz,
\end{equation}
where
\begin{equation}
E_{0}=\left(\frac{\hbar\omega_{0}}{\epsilon_{0}V}\right)^{1/2},
\end{equation}
\begin{equation}
B_{0}=\frac{\mu_{0}}{k}\left(\frac{\epsilon_{0}\hbar\omega_{0}^{3}}{V}\right)^{1/2},
\end{equation}
are sometimes referred to as the electric and magnetic fields ``per photon". In Eqs.~(\ref{eq:EL})
and ~(\ref{eq:MA}), the wavenumber
\begin{equation}
k=\frac{\omega_{0}}{c},
\end{equation}
where $c$ is the speed of light, and the optical field frequency is given by
\begin{equation}
\label{eq:ResFreq}
\omega_{0}=\frac{n\pi c}{L}, \,\,\ n=1,2,3,\ldots,
\end{equation}
which follows from the physical requirement that the length of the cavity has to be an integer number
of half wavelengths to support the electromagnetic mode. Using the fields of Eqs.~(\ref{eq:EL}) and
Eqs.~(\ref{eq:MA}) in Eq.~(\ref{eq:HamOP}) yields the harmonic oscillator Hamiltonian
\begin{equation}
\label{eq:HOPT}
H_{0}=\hbar\omega_{0}a^{\dagger}a.
\end{equation}
The variables $a$ and $a^{\dagger}$ correspond to destruction and creation operators for the photons
confined to the cavity, and
\begin{equation}
[a,a^{\dagger}]=1.
\end{equation}
Realistically, of course, the Fabry-Perot mirrors are not perfectly reflecting. In fact, some degree of
transmission is required so photons can be injected into the cavity, typically by a laser. To
account for this driving, the Hamiltonian of Eq.~(\ref{eq:HOPT}) is modified to \cite{AspelmeyerRMP2014}
\begin{equation}
\label{eq:HDrive}
H_{0}=\hbar\omega_{0}a^{\dagger}a+i\hbar F\left(a^{\dagger} e^{-i\omega_{l}t}-a e^{i\omega_{l}t}\right),
\end{equation}
where $\omega_{l}$ is the frequency of the driving laser, $F$ is related to the input laser power $P$
by
\begin{equation}
F=\sqrt{\frac{P\gamma_{0}}{\hbar\omega_{0}}},
\end{equation}
and $\gamma_{0}$ is the rate at which photons leak from the cavity.

The dynamical evolution of $a$ can be obtained from Eq.~(\ref{eq:HE}) (we presently ignore the effects
of $\gamma_{0}$ on this equation, but see Eq.~(\ref{eq:aEqn}) below)
\begin{equation}
\label{eq:HA}
\dot{a}=-i\omega_{0}a+Fe^{-i\omega_{l}t},
\end{equation}
which can be transformed using the substitution (which corresponds to entering the frame rotating at
the frequency of the input laser)
\begin{equation}
a\rightarrow a e^{i\omega_{l}t},
\end{equation}
to the equation
\begin{equation}
\label{eq:HAr}
\dot{a}=-i\Delta_{0}a+F,
\end{equation}
where
\begin{equation}
\Delta_{0}=\omega_{0}-\omega_{l},
\end{equation}
is the detuning of the laser from the cavity. The presence of the cavity mirror transmissivity also
introduces dissipation and noise into the cavity. The effect of these disturbances on the optical mode
can be accommodated using the Quantum Langevin approach and changes the time evolution of Eq.~(\ref{eq:HAr})
to \cite{AspelmeyerRMP2014}
\begin{equation}
\label{eq:aEqn}
\dot{a}=-i\Delta_{0}a-\frac{\gamma_{0}}{2}a+\sqrt{\gamma_{0}}a_{\mathrm{in}}+F,
\end{equation}
where $a_{\mathrm{in}}$ is the electromagnetic noise that comes in through the transmissive mirror,
and has zero mean and the correlator \cite{GenesPRA2008}
\begin{equation}
\langle a_{\mathrm{in}}(t)a_{\mathrm{in}}^{\dagger}(t')\rangle=\delta(t-t').
\end{equation}
at optical frequencies . The presence of the cavity decay $\gamma_{0}$ introduces a nonzero width to
the frequency $\omega_{0}$ of the electromagnetic field. With this inclusion of photon loss from the
cavity, the optical field is referred to as a \textit{quasi}-mode. Note that according to Eq.~(\ref{eq:aEqn})
the field variables $a$ and $a^{\dagger}$ each decay at the rate $\gamma_{0}/2$, while the oscillator energy
decays at $\gamma_{0}$. In the classical limit, the operator $a$ can be replaced by the complex-valued
dynamical variable $\alpha(t)$, which corresponds to the amplitude of the classical electromagnetic field
\cite{GerryBook}. In this limit, the number of photons in the cavity is $|\alpha(t)|^{2}.$

\subsection{Optomechanical cavity}
\label{subsec:OMCAV}
\begin{figure}
\begin{center}
\includegraphics[width=0.6\textwidth, height=0.2\textwidth]{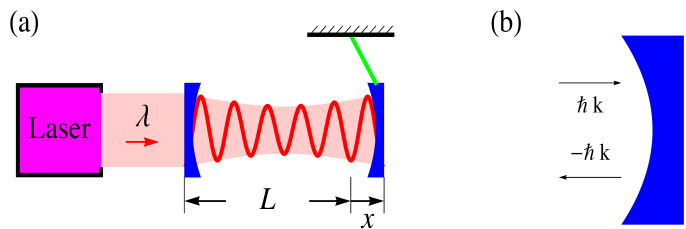}
\caption{(Color online) (a) An optomechanical cavity. The arrangement is the same as in Fig.~\ref{fig:P1},
but with one of the mirrors mounted as a harmonically oscillating pendulum with amplitude $x$ measured from
its equilibrium position. (b) Reversal of the photon momentum upon reflection from a perfectly reflecting
mirror.}
\label{fig:P2}
\end{center}
\end{figure}

We now consider a Fabry-Perot cavity where one mirror is fixed but the other is allowed to oscillate
harmonically along the cavity axis. This can be accomplished by suspending it like a pendulum, as in
Fig.~\ref{fig:P2}(a),
or by mounting it on a spring, for instance. We will derive the Hamiltonian for this system using heuristic semiclassical
arguments \cite{ShiAJP2013}; a rigorous quantum mechanical description can be found in \cite{LawPRA1995}. If the
mechanical oscillation frequency $\omega_{m}$ is much smaller than the spacing between two neighboring modes of
the cavity, then we can consider a single optical mode, since photon scattering into neighboring modes is then
negligible. Let us assume first that the two mirrors are both perfectly reflecting, and consider the reflection of a
single photon, as shown in Fig.~\ref{fig:P2}(b). The photon momentum before reflection is $\hbar k$ and after reflection
is $-\hbar k$. The net change of momentum of the photon is therefore $2\hbar k.$ The reflected photon returns
after being reflected from the stationary mirror, after a time $2L/c$. The change of momentum per unit time,
i.e. the force exerted by each photon is
\begin{equation}
\hbar g = \frac{2\hbar k}{2L/c},
\end{equation}
from which the optomechanical coupling constant may be written as
\begin{equation}
\label{eq:gOMVIB}
g=\frac{ck}{L},
\end{equation}
or using Eq.~(\ref{eq:ResFreq}) as
\begin{equation}
\label{eq:OMCoupling}
g=\frac{\omega_{0}}{L}.
\end{equation}
The total force due to all the photons present in the cavity is therefore $\hbar g a^{\dagger}a$. The work
extracted by the optical field in moving the mechanical oscillator from $L$ to $L+q$ yields the energy of
optomechanical interaction
\begin{equation}
\label{eq:Hint}
H_{\mathrm{int}}=-\hbar g a^{\dagger}a q,
\end{equation}
where all dynamical variables are quantum mechanical. The full optomechanical Hamiltonian then combines
Eqs.~(\ref{eq:HOPT}), (\ref{eq:HDrive}), and (\ref{eq:Hint})
\begin{equation}
H_{OM}=\frac{p^{2}}{2m}+\frac{1}{2}m\omega_{m}^{2}q^{2}+\hbar \omega_{0}a^{\dagger}a-\hbar g a^{\dagger}a q
+i\hbar F\left(a^{\dagger} e^{-i\omega_{l}t}-a e^{i\omega_{l}t}\right).
\end{equation}
In order to remove the explicit time-dependence, a transformation
\begin{equation}
U=e^{i\omega_{l}a^{\dagger}a t},
\end{equation}
into the frame rotating with the driving laser may be made, yielding,
\begin{equation}
H_{OM}'=U H_{OM}U^{\dagger}-i\hbar U\frac{\partial U^{\dagger}}{\partial t},
\end{equation}
or explicitly,
\begin{equation}
\label{eq:HOMDriveFrame}
H_{OM}'=\frac{p^{2}}{2m}+\frac{1}{2}m\omega_{m}^{2}q^{2}+\hbar \Delta_{0} a^{\dagger}a-\hbar g a^{\dagger}a q
+i\hbar F\left(a^{\dagger}-a\right).
\end{equation}
Sometimes it is convenient to use the variables of Eq.~(\ref{eq:qpTOaad}),
\begin{equation}
\label{eq:HOMP}
H_{OM}'=\hbar \Delta_{0} a^{\dagger}a+\hbar \omega_{m} b^{\dagger}b-\hbar g' a^{\dagger}a \left(b^{\dagger}+b\right)
+i\hbar F\left(a^{\dagger}-a\right),
\end{equation}
where
\begin{equation}
\label{eq:gp}
g'=g q_{0}.
\end{equation}
Accounting for noise and dissipation, the Quantum Langevin Equations for the optomechanical cavity are
\begin{equation}
\label{eq:QLEOPTO1}
\dot{q} = \frac{p}{m},
\end{equation}
\begin{equation}
\label{eq:QLEOPTO2}
\dot{p} = -m\omega_{m}^{2}q-\gamma_{m}p+\xi,
\end{equation}
\begin{equation}
\label{eq:QLEOPTO3}
\dot{a} =-\left(i\Delta_{0}+\frac{\gamma_{0}}{2}\right) a+ig'aq+F+\sqrt{\gamma_{0}}a_{\mathrm{in}}.
\end{equation}

\subsection{Optomechanical sensing and cooling of mechanical motion}
While the optomechanical cavity described above is associated with several physical phenomena, we
will briefly discuss only two below: displacement sensing and cooling of mechanical motion.
This abbreviated discussion will be sufficient for motivating the introduction of rotational
optomechanics below. For readers interested in other effects, such as bistability, squeezing, etc.,
we recommend several available reviews
\cite{VahalaReview2008,MarquardtReview2008,GirvinReview2009,MeystreReview2013,AspelmeyerRMP2014}.

Displacement sensing using optomechanical platforms is an application that relies on the fact
that a cavity photon picks up a phase that depends on the distance it travels, and hence on the displacement
of the mechanical oscillator in Fig.~\ref{fig:P2}. Mathematically, we can see from Eq.~(\ref{eq:QLEOPTO3})
that the time evolution of the optical mode $a$ depends on the oscillator displacement $q$. To
investigate further, let us assume the so-called `bad cavity' limit, where $\gamma_{0}$ is large and the
photons leave the cavity so quickly that the optical mode can rapidly adjust to instantaneous position of
the oscillator. In this regime, we can adiabatically eliminate the cavity mode, by setting $\dot{a}$ in
Eq.~(\ref{eq:QLEOPTO3}) to zero. This procedure yields
\begin{equation}
\label{eq:Fadia}
a=\frac{F}{\frac{\gamma_{0}}{2}+i\left(\Delta_{0}+g'q\right)},
\end{equation}
which shows the dependence of $a$ on $q$. Rewriting in polar form, and assuming $\Delta_{0}=0$ for
simplicity,
\begin{equation}
a (\Delta_{0}=0)=\frac{F}{\left[\left(\frac{\gamma_{0}}{2}\right)^{2}+\left(g'q\right)^{2}\right]^{1/2}}
e^{i\tan^{-1}\left(-\frac{g'q}{\gamma_{0}/2}\right)},
\end{equation}
which shows that the oscillator position information is contained in both the amplitude as well as
the phase of the cavity mode. In particular, the phase becomes more sensitive to $q$ if $g'/\gamma_{0}$ is
large, i.e. for stronger optomechanical coupling and low-loss cavities (although we are in the high loss
limit).
%By combining Eq.~(\ref{eq:Fadia}) with the input-output relation from quantum optics \cite{CollettPRA1984}
%\begin{equation}
%a_{\mathrm{out}}=a_{\mathrm{in}}-\sqrt{\gamma_{0}}a,
%\end{equation}
%we see that monitoring the phase of the output optical field provides continuous information about the
%oscillator position.
A more detailed analysis shows that in the `good cavity' limit $(\gamma_{0}$ small),
the displacement sensitivity of the device improves with the cavity finesse \cite{AspelmeyerRMP2014}. This
principle is the basis of interferometers such as LIGO which seek to transduce weak gravitational waves into
detectable mechanical displacements \cite{McClellandReview2011}.

Optical cooling of the mechanical motion is another application of the optomechanical cavity
\cite{AspelmeyerRMP2014}. Neglecting dissipation, noise and driving for simplicity, Eq.~(\ref{eq:HOMP}) reads
\begin{equation}
\label{eq:HCool}
H_{OM}'=\hbar \Delta_{0} a^{\dagger}a+\hbar \omega_{m} b^{\dagger}b-\hbar g' a^{\dagger}a \left(b^{\dagger}+b\right).
\end{equation}
The interaction term proportional to $a^{\dagger}a b$ corresponds to cooling of the mechanical motion, since
it describes a process in which a photon $(a)$ and a phonon $(b)$ are annihilated while a second photon
$(a^{\dagger})$ is created. Since energy has to be conserved during this process, the second photon must have
higher energy than the first, by exactly $\hbar \omega_{m}$, the phonon energy. This is allowed because the
photons belong to a quasi-mode with nonzero linewidth, and need not both have exactly the same frequency. The
key point is that the annihilation of a phonon cools the mechanical motion of the oscillator. Using similar
arguments, it can be shown that the interaction term proportional to $a^{\dagger}a b^{\dagger}$ corresponds
to oscillator heating. A detailed treatment indicates that by tuning the driving laser to below the cavity resonance
$(\Delta_{0} >0)$ allows the cooling to dominate over the heating; such a scheme has been used for laser cooling
a mesoscopic mechanical oscillator to its quantum ground state \cite{ChanNature2011}.

\section{Optical beams carrying angular momentum}
\label{sec:OAM}
In this section, we will provide a brief introduction to optical beams carrying angular momentum \cite{GalvezAJP2006}.
We will assume the reader is familiar with the fact that a photon can have spin angular momentum $\pm \hbar$,
corresponding classically to circularly polarized radiation. This is one way in which optical beams can carry
angular momentum. Another way is by encoding the appropriate structure in the beam phase \cite{OAMBook,KolobovBook,BekshaevBook,TorresBook,AndrewsBook,Andrews2Book,Allen1999,Padgett2004,Arnold2008,Yao2011}.
Each photon in the beam can then be thought of as carrying an OAM which can take unlimited values in
principle, and quite high values in practice. Although a clear separation of spin and orbital angular
momentum degrees of freedom is generally a subtle matter \cite{OAMBook,BekshaevBook}, in this article
we will operate under conditions where the requirements for effecting this separation are fulfilled.
Specifically, we will work in the paraxial regime of beam propagation, where the separation is valid.
Below we will introduce OAM-carrying beams by discussing a specific method for their generation. Based
on this discussion we will consider some general properties of such beams, which come in several families
\cite{Yao2011}. Subsequently, we will consider a particular family, namely the Laguerre-Gaussian modes,
for a more detailed exposition.

\subsection{Generation of OAM-carrying beams}
\label{sec:PWWAT}
\begin{figure}
\begin{center}
\includegraphics[width=0.8\textwidth, height=0.4\textwidth]{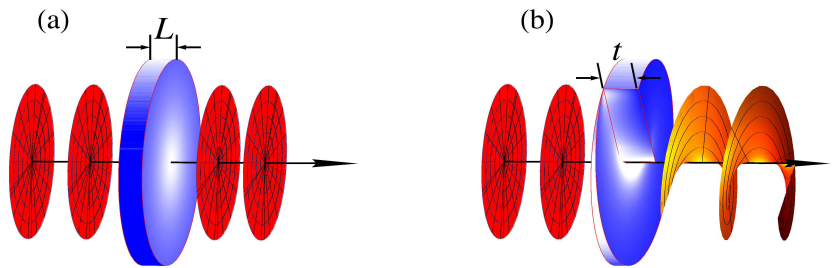}
\caption{(Color online) (a) Transmission of the wavefronts of a plane wave through a uniform slab of
thickness $L$ (b) Transmission of the wavefronts of a plane wave through a spiral phase plate of step height
$t$. The transmitted wavefronts in this case are helicoidal. Adapted with permission from \cite{Yao2011}.}
\label{fig:P3}
\end{center}
\end{figure}

There are several ways of generating beams carrying OAM, such as using lasers, lenses, holograms, spatial
light modulators and spiral phase plates \cite{AndrewsBook}. We will consider the relevant case of a
spiral phase plate, shown in Fig.~\ref{fig:P3}(b).
To understand what such a phase plate does to an optical beam, we first consider the simpler case of the
same beam passing through a uniform slab of thickness $L$, as shown in Fig.~\ref{fig:P3}(a). If the beam
incident on the slab is modeled as a plane wave of amplitude $E_{0}$, it will pick up a phase on transmission,
leading to an output field $E_{0}'e^{ikL},$ where $E_{0}'$ is the amplitude of the transmitted radiation,
$k$ is the wavevector of the optical beam, and $\hbar k$ is the linear momentum per photon. The
optical phase acquired by the beam therefore depends on the thickness $L$ of the slab.

When the same beam passes through the spiral phase plate, which has a uniform azimuthal gradient in its
thickness, it picks up a phase which depends on the angle $\phi$ about the light propagation direction,
\begin{equation}
\label{eq:OAMField}
E_{0}\rightarrow E_{0}'e^{il\phi},
\end{equation}
where
\begin{equation}
\label{eq:lint}
l= \frac{t}{\lambda},
\end{equation}
$t$ is the step height of the phase plate, and $\lambda$ the optical wavelength. Analogous to the linear
phase factor discussed above, where $\hbar k$ is the linear momentum, $l\hbar$ can be identified as the
orbital angular momentum carried by each photon in the transmitted beam. This identification can be verified by
rigorously calculating the angular momentum carried by the electromagnetic fields of the beam \cite{AllenPRA1992}.
By arranging for the plate thickness to be an integer multiple of the optical wavelength, we will restrict
ourselves to integer values of $l$ in Eq.~(\ref{eq:lint}). If $l$ is not an integer, the transmitted beam
consists of an infinite superposition of integer-$l'$ waves, whose average OAM is then the non-integer value
given by Eq.~(\ref{eq:lint}).

Although we have considered transmission through the spiral phase plate in our example above, reflected
photons pick up orbital angular momentum as well. In both cases the angular momentum is supplied by the spiral
phase plate. Usually the plate is clamped to an optical table and serves as a virtually infinite source or
sink of angular momentum. Lastly, we point out that the azimuthal phase factor $e^{il\phi}$ in the field
transmitted by the spiral phase plate is well-defined everywhere except on the $z$ axis, as $\phi$ is multiply
defined on that line. In a consistent physical picture of the field, the amplitude vanishes on the $z$-axis,
and the beam resultantly carries a singularity, namely a vortex. The intensity profile of the beam therefore
presents a node at the core of the vortex, i.e. on the $z$ axis, see Fig.~\ref{fig:P4} below. The OAM $l$,
which quantifies the beam vorticity, is sometimes referred to as the `topological charge' of the optical
beam.

We have presented an idealized description thus far, using a plane wave, which has infinite transverse extent.
In practice, the beam has a transverse profile which decreases in intensity away from the propagation axis.
We now move on to the discussion of such modes.

\subsection{Laguerre-Gaussian modes} An appropriate expression for an OAM-carrying field is
\begin{equation}
E\left(\textbf{r}\right)= u\left(\textbf{r}\right)e^{il\phi},
\end{equation}
which accounts for the transverse coordinate dependence of the field. Fields of this type form solutions to
the paraxial wave equation in free space and are therefore bonafide modes of the electromagnetic field in that
approximation \cite{Andrews2Book}. There are several families of beams that carry OAM, corresponding to different
forms for $u\left(\textbf{r}\right)$, including Bessel, Mathieu, and Hypergeometric-Gaussian modes \cite{Yao2011}.
We consider the Laguerre-Gaussian fields for which
\begin{equation}
E\left(\textbf{r}\right)= u_{lp}\left(\textbf{r}\right)e^{il\phi},
\end{equation}
and the normalized mode function is given by \cite{Andrews2Book}
\begin{equation}
\label{eq:LGmode}
\fl u_{lp}\left(\textbf{r}\right)=\frac{C_{lp}}{\sqrt{w(z)}}\left[\frac{\rho\sqrt{2}}{w(z)}\right]^{|l|}
e^{-\frac{\rho^{2}}{w^{2}(z)}} L_{p}^{|l|}\left[\frac{2\rho^{2}}{w^{2}(z)}\right]
e^{\frac{-ik\rho^{2}z}{2\left(z_{R}^{2}+z^{2}\right)}}e^{il\phi}
e^{i\left(2p+|l|+1 \right)\tan^{-1}\left(\frac{z}{z_{R}}\right)},
\label{eq1}
\end{equation}
where the normalization 
\begin{equation}
C_{lp}=\sqrt{\frac{2^{|l|+1}p!}{\pi\left(p+|l|\right)}},
\end{equation}
the index $l$ corresponds to the number of azimuthal nodes, the index $p$ determines the number of radial
nodes \cite{KarimiPRA2014}, $w(z)$ is the beam waist, $L_p^{|l|}$ is the asssociated Laguerre polynomial,
and $z_{R}=\pi w^{2}(0)/\lambda$ is the Rayleigh range, $\lambda$ being the optical wavelength. In 
Eq.~(\ref{eq:LGmode}), $l$ can be either a positive or negative integer, while $p$ can only assume positive 
integer values.

\begin{figure}
\begin{center}
\includegraphics[width=0.5\textwidth]{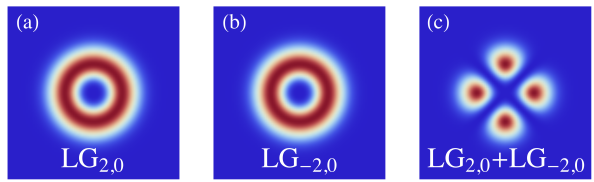}
\caption{(Color online) (a) Intensity pattern for an $l=2,p=0$ Laguerre-Gaussian beam (b) Intensity pattern
for an $l=-2,p=0$ Laguerre-Gaussian beam (c) Intensity pattern for a superposition of $l=2, p=0$ and $l=-2, p=0$
Laguerre-Gaussian beams.}
\label{fig:P4}
\end{center}
\end{figure}

The intensity profile of a Laguerre-Gaussian mode with $l\neq 0$ has a node at the center, and appears shaped as
an annulus, see Fig.~\ref{fig:P4}(a) for $l=2$. The interference pattern between two LG$_{\pm l}$ beams consists
of $2l$ lobes arranged in a circle, see Fig.~\ref{fig:P4}(c) for the case where $l=2.$
With this brief introduction to OAM-carrying beams, we are now ready to combine mechanical motion with photonic
angular momentum.

\section{Torsional cavity optomechanics}
\label{sec:TorsionalOpto}
In this section, we will consider some optomechanical platforms that couple photonic angular momentum to
mechanical torsional motion. We will first introduce a straightforward analog of the linear optomechanical cavity
presented in Section~\ref{subsec:OMCAV}, and then consider a variation on the same theme.

\subsection{Spiral Resonator}
\label{subsec:SR}
\begin{figure}
\begin{center}
\includegraphics[width=0.6\textwidth, height=0.4\textwidth]{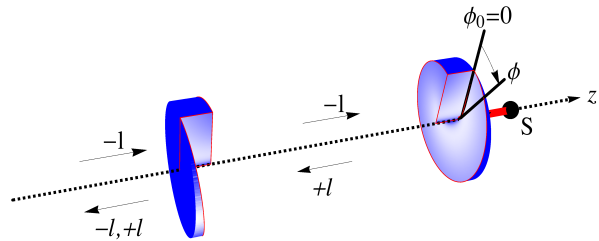}
\caption{(Color online) A picture of the spiral resonator discussed in Section~\ref{subsec:SR}. Adapted with
permission from \cite{MBPRL2007}.}
\label{fig:P5}
\end{center}
\end{figure}

We begin with a torsional analog to the standard linear vibrational optomechanical cavity of
Section~\ref{subsec:OMCAV} \cite{MBPRL2007}, as shown in Fig.~\ref{fig:P5}. In this analogy, the spherical
mirrors used in a Fabry-Perot are replaced by spiral phase plates described in Section~\ref{sec:PWWAT}. One
of the plates is fixed in place, while the other  oscillates harmonically about the cavity axis, thus executing
torsional motion. The angular displacement of the moving plate is measured by the angle $\phi$. A beam with OAM
$l\hbar$ is incident on the cavity.  The spiral phase plates reverse the linear momentum as well as the OAM of
the cavity photons upon reflection. This reversal of OAM can be arranged by selecting the step heights and
handedness on the spiral plates so that one of them adds, and the other subtracts, the same OAM from the cavity
photons, as shown in Fig.~\ref{fig:P5}.
The optomechanical coupling can be derived analogously to Section~\ref{subsec:OMCAV}  if we assume that
the two spiral phase plates are perfectly reflecting. In this case, the moving plate reverses the photonic
OAM from $l\hbar$ to $-\l\hbar$ upon reflection, resulting in a momentum change of $2l\hbar$ per photon.
This happens once every cavity round trip time $2L/c$, if the cavity length is $L$. The change of angular
momentum per unit time, i.e. the torque exerted by each photon is given by
\begin{equation}
\hbar g_{\phi} = \frac{2\hbar l}{2L/c},
\end{equation}
which may be compared to Eq.~(\ref{eq:OMCoupling}) and from which the \textit{optotorsional} coupling constant,
may be obtained as
\begin{equation}
\label{eq:gOMTOR}
g_{\phi}=\frac{cl}{L},
\end{equation}
which is analogous to Eq.~(\ref{eq:gOMVIB}). The total torque due to all the photons present in the cavity
is $\hbar g_{\phi} a^{\dagger}a$. The work extracted from the torsional oscillator in deflecting
it by an angle $\phi$ yields the energy of optomechanical
interaction
\begin{equation}
\label{eq:HintT}
H_{\mathrm{int}}^{\phi}=-\hbar g_{\phi} a^{\dagger}a \phi,
\end{equation}
The entire optomechanical Hamiltonian may now be assembled in analogy to Eq.~(\ref{eq:HOMDriveFrame})
\begin{equation}
\label{eq:HOMT}
H_{OM}'^{\phi}=\frac{L_{z}^{2}}{2I}+\frac{1}{2}I\omega_{\phi}^{2}\phi^{2}+\hbar \Delta_{0} a^{\dagger}a
-\hbar g_{\phi} a^{\dagger}a \phi+i\hbar F\left(a^{\dagger}-a\right),
\end{equation}
where $L_{z}$ is the angular momentum of the oscillating spiral phase plate about the cavity axis, $I$ is its moment
of inertia, and $\omega_{\phi}$ is the frequency of torsional oscillation. The first two terms describe the torsional
energy of oscillation, the third term the optical mode energy, the fourth term the optomechanical interaction, and
the fifth term the effect of external driving on the cavity. If the moving spiral phase plate has mass $M$ and radius
$R$, we may treat it as a disk and evaluate the moment of inertia
\begin{equation}
I=\frac{MR^{2}}{2},
\end{equation}
about an axis passing through its center. In Eq.~(\ref{eq:HOMT}), all dynamical variables are quantum mechanical.
Specifically, the torsional variables obey the commutation relation
\begin{equation}
\left[L_{z},\phi\right]=-i\hbar.
\end{equation}
Since Eq.~(\ref{eq:HOMT}) has the same form as Eq.~(\ref{eq:HOMDriveFrame}), we may expect to implement the same
optomechanical physics on this new platform, i.e. detection of torsional displacements as well as  cooling of
torsional motion should be possible. Interestingly, the optotorsional coupling in (\ref{eq:gOMTOR}) depends on $l$,
which may be adjusted by the experimentalist. In recent experiments, upto $l=300$ has been used
\cite{FicklerScience2012}.

Unlike the standard Fabry-Perot, which has been investigated for more than a century, the spiral phase plate
resonator appeared in the theoretical literature only once (without the mechanical degree of freedom)
\cite{Kudryashov2000} before the optomechanical version was proposed. Therefore, much fundamental information
about the spiral resonator remains to be discovered. As a first step in this direction, our group has recently
analyzed the ray optics of this resonator \cite{Eggleston2013}; related work has also recently been carried
out for low finesse spiral etalons \cite{RumalaAO2015}. No experimental realization of the spiral phase
resonator seems to exist, possibly due to the low reflectivity $(\leq 95\%)$ and high diffractive losses
presented by currently available spiral phase plates. These limitations preclude the implementation of efficient
displacement sensing or torsional cooling, both of which require high-finesse cavities. Interestingly, a design
equivalent to a spiral phase resonator (but without any mechanical degrees of freedom) has been implemented in
a situation where the mirror losses are offset by the presence of a gain medium - i.e. in a laser
\cite{OronProgOpt2001}.

\subsection{Nanorods and Windmills}
\label{subsec:NW}
In this section we will consider the proposal of Romero-Isart et al. \cite{IsartNJP2010} for coupling a rod
to OAM-carrying cavity modes, and its subsequent generalization to nano-windmills \cite{Shi2013}. These proposals
rely on the ability of light to optically trap dielectric particles at the intensity maxima of an optical beam.
The trapping may be understood intuitively if the dielectric particle is smaller than an optical wavelength
and thus may be assumed to behave as a point dipole. If the dipole moment is $\vec{d}$, the energy of interaction
with an optical beam of electric field $\vec{E}$ is
\begin{equation}
\label{eq:HintTrap}
H_{t}=-\vec{d}\cdot\vec{E}.
\end{equation}
If the particle is a neutral but linearly polarizable dielectric, with polarizability $\alpha$, then the
dipole moment is an induced one, and is given by
\begin{equation}
\label{eq:Dipole}
\vec{d}=\alpha \vec{E}.
\end{equation}
Combining Eqs.~(\ref{eq:HintTrap}) and (\ref{eq:Dipole}), we obtain
\begin{equation}
\label{eq:HAMDip}
H_{t}= -\alpha \left|\vec{E}\right|^{2}.
\end{equation}
Now since $\left|\vec{E}\right|^{2}=2I_{0}/c\epsilon_{0}$ where $I_{0}$ is the light intensity, clearly $H_{t}$
is a minimum when the particle is located at an intensity maximum of the beam. At such maxima, in fact, the
force from the optical beam can levitate the particle in space against the effect of gravity. In practice,
it is quite common to use the focus of a Gaussian beam as an optical trap for the particle \cite{Padgett1997}.
To lowest order in the particle displacement, such a trap provides harmonic confinement in all three directions
in space.

\begin{figure}
\begin{center}
\includegraphics[width=0.5\textwidth, height=0.15\textwidth]{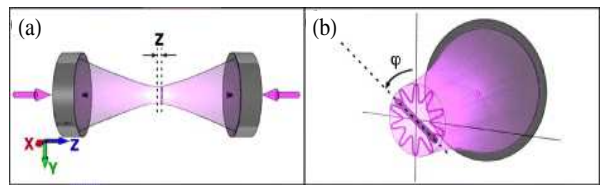}
\caption{(Color online) The system discussed in Section~\ref{subsec:NW}. (a) shows the trapping configuration
along the cavity axis (b) shows the angular optical lattice in which the rod is trapped and executes torsional
motion about the cavity axis. Adapted with permission from \cite{IsartNJP2010}.}
\label{fig:P6}
\end{center}
\end{figure}

In their proposal, Romero-Isart suggest trapping a nanorod with multiple Laguerre-Gaussian cavity
modes, in such a manner that the rod executes torsional oscillations about the cavity axis, see Fig.~\ref{fig:P6}.
The trapping field is considered to be an experimentally changeable parameter and does not possess dynamics
of its own. An additional Laguerre-Gaussian mode with $p=0$ is introduced to detect and cool the torsional
oscillations. The corresponding Hamiltonian is essentially of the same form as in Eq.~(\ref{eq:HOMT}), and
therefore we do not repeat its full derivation here. We only note that the optomechanical coupling in the
case of the rod was calculated using the Bethe-Schwinger perturbation theory formula \cite{NovotnyBook}
\begin{equation}
\label{eq:FreqMode}
\frac{\omega_{c}(\phi)}{ \omega_{c}(\phi_{0})}\simeq 1-\frac{\int_{V} (\epsilon-1)\left|u_{l0}({\bf r})\right|^{2}d {\bf r}}
{2\int_{V'} \left|u_{l0}({\bf r})\right|^{2}d {\bf r}},
\end{equation}
which accounts for the variation of the cavity resonance $\omega_{c}(\phi)$ due to the change in the angular position
of the rod $\phi$ away from its equilibrium value $\omega_{\phi_{0}}$ and where $\epsilon$ and $V$ are the
dielectric constant and volume, respectively, of the windmill, and $V'$ is the volume of the optical cavity. The
optomechanical coupling can then be found as
\begin{equation}
g_{l,0}=\sqrt{\frac{\hbar}{I \omega_\phi}}\frac{\partial \omega_{c}(\phi)}{\partial \phi |_{\phi_{0}}},
\end{equation}
from Eq.~(\ref{eq:FreqMode}). The Bethe-Schwinger formula is applicable when the dimensions of the rod are
small enough that its presence in the cavity shifts the optical resonance only by a small amount. Interestingly,
the authors point out that a suitable nanorod could be the Tobacco Mosaic Virus, thus opening up a way to observe
quantum effects in living organisms.

\begin{figure}
\begin{center}
\includegraphics[width=0.6\textwidth]{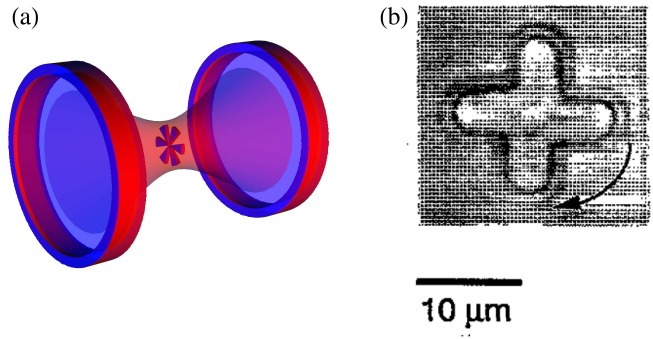}
\caption{(Color online) (a) A nanowindmill levitated inside an optical cavity. Adapted with permission
from \cite{Shi2013} (b) A photomicrograph of a microfabricated windmill. Reproduced with permission from
\cite{HigurashiAPL1994}.}
\label{fig:P7}
\end{center}
\end{figure}

The work of Romero-Isart et al., has been generalized further by considering, instead of a single rod, a windmill-shaped
dielectric, which may be thought of as a collection of rods, see Fig.~\ref{fig:P7}(a) \cite{Shi2013}.
The advantage of a windmill shape is that it provides larger overlap with the cavity OAM modes than a single rod,
and hence higher optomechanical coupling. We note that micromachined windmills are available in the laboratory,
and have been held and rotated in optical tweezers using radiation pressure from optical beams \cite{HigurashiAPL1994}
, see Fig.~\ref{fig:P7}(b). While Romero-Isart considered only $p=0$ beams [see Eq.~(\ref{eq:LGmode})], it was later
found that using modes with nonzero $p$ yields better optomechanical coupling, since such modes match the shape of
the windmill better \cite{Shi2013}.

A recent experimental development along these lines was the demonstration of the transit
of rotating nanorods through an optical cavity excited by a Gaussian optical beam \cite{KuhnNL2015}. More
sophisticated theoretical proposals also exist, incorporating multiple nanodumbbells into an optical cavity
and predicting optically induced ordering and nematic transitions analogous to those observed in liquid
crystals \cite{LechnerPRL2013}. Torsional optomechanics realized by coupling wave guide deformations with
circularly polarized photons has also been proposed \cite{ZhangOE2011}. Thus far we have considered the
coupling of optical angular momentum to mechanical torsional motion. We now proceed to examine the coupling
of optical angular momentum with free mechanical rotation.

\section{Rotational cavity optomechanics}
\label{sec:RCO}
In this section we introduce the coupling of optical angular momentum with unhindered mechanical rotation.
First we consider a particle rotating in the evanescent field of an optical resonator, and then a
configuration in which the particle experiences the intracavity field.

\subsection{Evanescently trapped nanoparticle}
\label{subsec:NanoRotout}
\begin{figure}
\begin{center}
\includegraphics[width=0.7\textwidth]{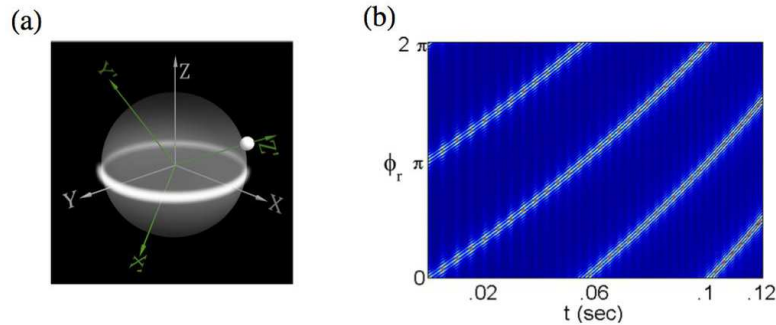}
\caption{(Color online) (a) The system described in Section~\ref{subsec:NanoRotout}. The large circle denotes a
microsphere, the bright band along the equator is a whispering gallery mode which confines the optical field,
and the small circle is a nanoparticle trapped in orbit around the sphere (b) Surface field intensity at the
microsphere equator as a function of time. Reproduced with permission from \cite{DeychPRA2011}.}
\label{fig:P8}
\end{center}
\end{figure}

In this section, we consider a subwavelength dielectric particle captured in orbital motion around a spherical
whispering gallery mode resonator, as analyzed by Rubin et al. \cite{DeychPRA2011,ArnoldOE2009}, and shown in
Fig.~\ref{fig:P8} (a).
To our knowledge, this was the first configuration in which fully rotational cavity optomechanics was studied
\cite{CheungPRA2012}. The resonator itself is a dielectric microsphere. The optical mode is confined by total
internal reflection near the surface of the resonator; the authors specifically consider whispering gallery
modes that are localized near the resonator equator, as in Fig.~\ref{fig:P8}. Some of the light escapes from
the spherical resonator in the form of an evanescent field. It is this light that is responsible for trapping
the subwavelength dielectric particle `and driving its motion. The optomechanical coupling in this case is
somewhat complicated and will not be derived here. The complications arise because the particle stays quite
close to the dielectric sphere and therefore modifies the charge distribution on the resonator, effects which
are negligible in the optomechanical systems described thus far.

The resulting particle dynamics are also quite involved. The radial, polar and azimuthal motions of the particle
are driven by the radiation field, and are coupled to each other. In particular, the orbital rotation
of the particle about the spherical resonator is driven by the optical torque, which turns out to be
nonconservative. However, as in standard optomechanics, the particle motion acts back on the field and
therefore can be deciphered by monitoring, for example, the field intensity at the resonator surface.
Figure ~\ref{fig:P8}(b) shows the surface field intensity along the resonator equator. The intensity variations
along the horizontal axis represent radial particle oscillations, and from their non-uniform spacing the angular
azimuthal acceleration of the particle can be found. Similarly, the intensity variations along the vertical axis
correspond to polar oscillations. While this configuration displays many interesting features, the role played by
electromagnetic angular momentum is somewhat limited. Assuming dipolar scattering of photons, the authors show that
the nanoparticle couples only to three optical angular momenta, namely $0$ and $\pm h$.

\subsection{Nanoparticle rotating inside a cavity}
\label{subsec:NanoRot}
\begin{figure}
\begin{center}
\includegraphics[width=0.3\textwidth]{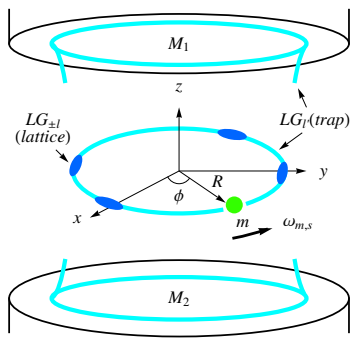}
\caption{(Color online) The system referred to in Section~\ref{subsec:NanoRot}. A nanoparticle of mass $m$ rotates
around an optical ring potential and is probed by an angular optical lattice inside a cavity. Reproduced with
permission from \cite{MBJOSA2015}.}
\label{fig:P9}
\end{center}
\end{figure}

In this section we will analyze a rotating nanoparticle located \textit{inside}
an optical cavity, as shown in Fig.~\ref{fig:P9} \cite{MBJOSA2015}.
Unlike in the configuration of Section~\ref{subsec:NanoRotout}, the usual light-matter coupling applies. This is
because the rotating particle is far away from the resonator boundaries. This configuration is the free
rotational analog of the linear optomechanical system of Section~\ref{subsec:OMCAV}, and we will examine
it in some detail. Our exposition below will be aimed at establishing the possibility of accurate rotation
sensing, i.e. implementing velocimetry of the particle, using a cavity. We will restrict ourselves to a classical
analysis. A quantum analysis of rotation is expected to be more subtle \cite{CarruthersRMP1968},

The optical resonator in this case is a Fabry-Perot, with both mirrors fixed in place. The cavity is excited
by a trap beam, which is a Laguerre-Gaussian mode with OAM $l_{t}=2$. This mode presents an intensity maximum on
a ring [see Fig.~\ref{fig:P9} and also Fig.~\ref{fig:P4}(b)] of radius
\begin{equation}
R=w_{0}\left(\frac{\left|l_{t}\right|}{2}\right)^{1/2},
\end{equation}
at the center of the cavity. A dielectric particle of diameter smaller than the optical wavelength is trapped
on this ring via the optical gradient force, see Eqs~(\ref{eq:HAMDip}). We consider this particle to be rotating
on this ring at a uniform angular velocity $\omega_{m,s}$ set by the balance of an external torque
and the damping produced by the background gas particles as realized experimentally \cite{SacconiOL2001}. We
assume conditions where the coupling to radial and polar motion are negligible, i.e. tight confinement on
the ring, and do not consider these degrees of freedom any further. The trap mode is treated as a parametric
quantity, without any associated dynamics.

An additional optical field, a superposition of two degenerate counter-rotating Laguerre-Gaussian modes
with OAM $+l$ and $-l$ respectively, is introduced to probe the rotational motion of the nanoparticle.
Near the trapping ring, the modes set up an optical lattice with $2l$ maxima, described by the function
\begin{equation}
\label{eq:ProbeMode}
\left|u_{l,0}\right|^{2}=\frac{1}{\left|l\right|!}\left(\frac{R\sqrt{2}}{w_{0}}\right)^{\left|l\right|}
e^{-\frac{2R^{2}}{w_{0}^{2}}}\cos^{2}\left(k_{p}z\right)\cos^{2}\left(\phi\right),
\end{equation}
where $k_{p}$ is the probe wave-vector. Figure ~\ref{fig:P4}(c) and Fig.~\ref{fig:P9} show the lattice for
$l=2$. Since the particle dimensions are smaller than the cavity length $L$ and the optical wavelength
$\lambda$, Eq.~(\ref{eq:FreqMode}) can be used to find the optorotational coupling $g(l)$, using
Eq.~(\ref{eq:ProbeMode}) as the input. This yields
\begin{equation}
\omega_{c}(\phi)=\omega_{c}-g(l)\cos^{2}l\phi,
\end{equation}
with
\begin{equation}
g(l)= \left(\epsilon_{r}-1\right)\frac{2^{\frac{l+3}{2}}}{\Gamma\left(\frac{l+1}{2}\right)}
\left(\frac{R}{w_{0}}\right)^{l}\frac{V}{\pi w_{0}^{2}L}e^{-\frac{2R^{2}}{w_{0}^{2}}},
\end{equation}
where $\epsilon_{r}$ is the relative dielectric constant of the nanoparticle and $\Gamma$ is a Gamma
function.

The dynamics of the optical probe field can be described using the variable $a(t)$ such that
$\left|a(t)\right|^{2}$ is the number of probe photons inside the cavity at any time $t$, see the
very end of Section \ref{subsec:CavQ}. Choosing the cavity axis as the $z$ direction, the angular
momentum of the particle is $L_{z}$. To describe the rotational coordinate of the particle we use
the periodic variable
\begin{equation}
U_{l}=e^{2il\phi},
\end{equation}
where $\phi$ is the instantaneous angular displacement of the particle, see Fig.~\ref{fig:P9}.

The coupled equations of motion in the frame of the laser driving the cavity with the $LG_{\pm}$ fields
can be written in analogy to Eqs.~(\ref{eq:QLEOPTO1})-(\ref{eq:QLEOPTO3})
\begin{eqnarray}
\label{eq:AngleEvolution}
\dot{U_{l}}&=&\frac{i 2 l U_{l} L_{z}}{I},\\
\label{eq:AngularMomentumEvolution}
\dot{L_{z}}&=&-\gamma_{m}L_{z}-2il\hbar g(l)\left(U_{l}-U_{l}^{*}\right)\left|a(t)\right|^{2}+\tau+\tau_{in}.\\
\label{eq:OpticalFieldEvolution}
\dot{a}&=&\left\{i\left[\Delta'-\frac{g(l)}{2}\left(U_{l}+U_{l}^{*}\right)\right]
-\frac{\gamma_{0}}{2}\right\}a+\sqrt{\gamma}a_{in},\\
\nonumber
\end{eqnarray}
where
\begin{equation}
\Delta'=\Delta-\frac{g(l)}{2},
\end{equation}
$\Delta=\omega_{d}-\omega_{c}$ being the detuning between the $LG_{\pm l}$ driving laser at frequency $\omega_{d}$
and $\omega_{c}$, $\gamma_{0}$ is the cavity loss rate, $I=m R^{2}$ is the dielectric particle's moment of inertia
about the axis of rotation and $a_{in}=\sqrt{P_{in}/\hbar\omega_{c}}$, where $P_{in}$ is
the input probe power. Classical laser noise and the thermal contribution to the radiation mode have been assumed to
be negligible. The rotor mechanical damping is $\gamma_{m}$, and $\tau_{in}$ is a Langevin torque with
zero mean and the two-time fluctuation correlation (analogous to Eq.~\ref{eq:harmcorr2})
\begin{equation}
\langle \delta \tau_{in}(t)\delta \tau_{in}(t')\rangle=2I \gamma_{m}k_{B}T\delta(t-t'),
\end{equation}
signifying white noise and responsible for Brownian motion. Generally, we will assume that the externally applied
torque is larger than the torque due to the optical lattice,
i.e. $\tau > 2il \hbar g(l)\left(U_{l}-U_{l}^{*}\right)\left|a(t)\right|^{2}.$

\begin{figure}
\begin{center}
\includegraphics[width=0.4\textwidth]{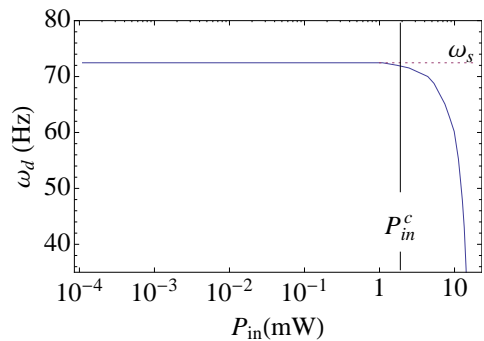}
\caption{(Color online) The optically detected mechanical rotation frequency $\omega_{d}$ as a function of probe
lattice power $P_{\mathrm{in}}$ in the setup of Fig.~(\ref{fig:P9}). The detected frequency accurately mirrors
the rate $\omega_{s}$ [Eq.~(\ref{eq:SidebandFrequency})] in the regime of linear response, i.e. for probe
power less than $P^{c}_{\mathrm{in}}$, but is affected for higher powers, degrading the accuracy of the measurement.
Adapted with permission from \cite{MBJOSA2015}.}
\label{fig:P10}
\end{center}
\end{figure}

Comparing Eq.~(\ref{eq:OpticalFieldEvolution}) to Eq.~(\ref{eq:QLEOPTO3}) shows that the phase of the optical field is
sensitive to the $(U_{l}+U_{l}^{*})/2=\cos 2 l\phi$ quadrature of the mechanical rotation, and that this
sensitivity improves linearly with the cavity finesse, and may therefore be used to accurately detect the
particle rotation. We elaborate on this procedure below. The steady-state of the system can be found readily by
equating the time derivatives in Eqs.~(\ref{eq:OpticalFieldEvolution})-(\ref{eq:AngularMomentumEvolution}) to zero,
\begin{equation}
\label{eq:SteadyState}
\omega_{m,s}=\frac{L_{z,s}}{I}=\frac{\tau}{I\gamma_{m}},
\end{equation}
\begin{equation}
\label{eq:SteadyState2}
U_{l,s}=0,
\end{equation}
\begin{equation}
\label{eq:SteadyState3}
a_{s}=\frac{\sqrt{\gamma}a_{in}}{\left[\Delta'^{2}+\left(\frac{\gamma_{0}}{2}\right)^{2}\right]^{1/2}}.
\end{equation}
We note from Equation ~(\ref{eq:SteadyState}) that the torque due to the probe lattice makes no
contribution, on average, to the rotation rate $\omega_{m,s}$. We can examine the linear response
of the system (i.e. the response of the system to small changes away from the steady state) by
considering each variable as the sum of a mean value and a small fluctuation, e.g.
\begin{equation}
a=a_{s}+\delta a.
\end{equation}
For low probe powers, we find
\begin{eqnarray}
\label{eq:linearresponse1}
\dot{\delta a}&=&-i \frac{g(l)}{2}a_{s}\left(\delta U_{l}+\delta U_{l}^{*}\right)
+\left(i\Delta'-\frac{\gamma}{2}\right)\delta a,\nonumber \\
\label{eq:linearresponse2}
\dot{\delta U_{l}}&=&\frac{i 2 l}{I}L_{z,s}\delta U_{l},\\
\label{eq:linearresponse3}
\dot{\delta L_{z}}&=&-\gamma_{m}\delta L_{z}+2il \hbar g(l)\left|a_{s}\right|^{2}\left(\delta U_{l}-\delta U_{l}^{*}\right)+\delta\tau_{in}.\nonumber\\
\nonumber
\end{eqnarray}
Solving this set of equations given the initial conditions $\delta a(0),\delta U_{l}(0)$ and $\delta L_{z}(0)$,
we find on resonance $(\Delta'=0)$, and ignoring cavity losses $(\gamma_{0}=0)$, that the Fourier transform of
$\delta a(t)$ is
\begin{equation}
\label{eq:SidebandSpectrum1}
\delta a\left(\omega\right)= A \delta\left(\omega\right)
+B^{*}\delta\left(\omega-\omega_{s}\right)-B\delta\left(\omega+\omega_{s}\right),
\end{equation}
where
\begin{equation}
A=\sqrt{2\pi}\delta a(0)+B-B^{*},
\end{equation}
\begin{equation}
B=\frac{\sqrt{2\pi}a_{s}g(l)\delta U_{l}(0)}{\omega_{m,s}},
\end{equation}
and
\begin{equation}
\label{eq:SidebandFrequency}
\omega_{s}=2l\omega_{m,s}.
\end{equation}
The Dirac delta at $\omega=0$ on the right hand side of Eq.~(\ref{eq:SidebandSpectrum1}) corresponds to the carrier
optical frequency, and the Dirac deltas at $\omega=\pm \omega_{s}$ to sidebands which are created as the particle
rotating at the frequency $\omega_{m,s}$ encounters $2l$ optical lattice sites. These sidebands fundamentally arise
from the rotational Doppler shift \cite{BirulaPRL1997,BarreiroPRL2006} imprinted on the cavity photons by the
mechanical motion. The rotation frequency peak at $\omega_{s}$ can thus be recovered by homodyning (i.e. interfering
with a piece which was split off from the probe beam before it enters the cavity) the probe beam $LG_{\pm l}$
once it has exited the cavity \cite{AspelmeyerRMP2014}. An investigation into the nonlinear regime of high probe
power shows that the measurement accuracy eventually degrades as shown in Fig.~\ref{fig:P10}.
At very high powers, the nanoparticle is simply trapped in one of probe lattice wells.

While the description above provides some initial exploration of the coupling between optical and particle
OAM, much remains to be accomplished in this area. Particularly, the quantum regime needs to be explored.
A situation where the quantum regime could be relevant is when the dielectric nanoparticle is replaced by
a single atom. No experiment to date has been able to detect or manipulate the center-of-mass angular momentum of
a single atom, see Section~\ref{subsec:RCLO} below. This is in contrast to other degrees of freedom, such as
electronic and translational center-of-mass for which fine control has been established at the single atom level
\cite{MeschedeAMOP2006,WinelandRMP2013,WiemanRMP2006}. Such control is desirable for rotational degrees of
freedom from a fundamental perspective, but also for applications such as quantum information processing
\cite{Zeilinger2012,MirhosseiniNJP2015}.

\subsection{Surface Acoustic Waves}
\label{subsubsec:SAW}
\begin{figure}
\begin{center}
\includegraphics[width=0.4\textwidth]{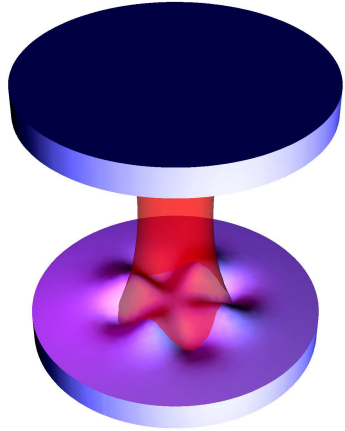}
\caption{(Color online) The system described in Section~\ref{subsubsec:SAW}. Surface acoustic waves
on the lower mirror couple to the optical mode supported by the cavity. Reproduced with permission from
\cite{ShiJPhysB2013}.}
\label{fig:P11}
\end{center}
\end{figure}

In this section we will consider the coupling of Laguerre-Gaussian optical cavity modes with acoustic
waves on the surface of a highly reflective mirror, see Fig.~\ref{fig:P11}.
This mirror is one of two that make up
a Fabry-Perot, of the kind described in Section~\ref{subsec:OMCAV}. The acoustic excitations manifest
as vibrations of the mirror surface along the cavity axis. Importantly, the vibrational modes also carry
azimuthal structure. To our knowledge, Briant et al. were the first to experimentally demonstrate that
standard optomechanical physics of Eq.~(\ref{eq:HOMP}) could be realized by such vibrational modes,
which do not involve any global motion of the mirror \cite{Briant2003}. They were able to detect the
azimuthal profile of each mode by scanning a sharply focused Gaussian beam, whose waist was much smaller
than that of the surface acoustic mode, across the mirror surface and following the resulting phase shifts
in the reflected cavity beam.

The experimentally acquired mechanical mode vibrational frequencies and spatial mode profiles could be fit
well by an analytical theory in the limit where the radius of curvature of the mirror was much larger than
its thickness. In this limit, the elasticity equation, which determines the mechanical modes, took a form
identical to the \textit{optical} paraxial equation. This implied that each mechanical mode could be
characterized as a Laguerre-Gaussian mode, as confirmed by the experimental data. In this case the mechanical
phonons of the mirror surface acoustic field may be thought of as harmonic oscillator quanta which carry OAM
transverse to their direction of oscillation. They are therefore analogous to
Laguerre-Gauss photons of the electromagnetic field which are also quanta of a harmonic oscillator,
and also carry OAM in the same manner.

The work of Briant et al. was extended in ~\cite{ShiJPhysB2013} which subsequently explored the coupling
of the mentioned mechanical Laguerre-Gaussian modes (with indices $l'$ and $p'$) to optical cavity (with indices
$l$ and $p$) Laguerre-Gaussian modes. The interaction Hamiltonian in this case is found by expanding the acoustic
as well as optical fields in terms of Laguerre-Gaussian modes [analogous to Eqs.~(\ref{eq:EL}) and ~(\ref{eq:MA})
which expand the field in terms of sines and cosines], and yields
\begin{equation}
\label{eq:SAW}
H_{\mathrm{SAW}} = \hbar g_{\mathrm{SAW}}a^{\dagger}a\left(c+c^{\dagger}\right),
\end{equation}
which is exactly of the same form as Eq.~(\ref{eq:Hint}), except for the coupling constant
\begin{equation}
\label{eq:gSAW}
g_{\mathrm{SAW}}=g'\delta_{\left|l'\right|,2\left|l\right|}\xi_{lpp'},
\end{equation}
where $g'$ is given in Eq.~(\ref{eq:gp}) and the full expression for $\xi_{lpp'}$ has not been reproduced
here as it is quite lengthy. From the Kronecker delta in Eq.~(\ref{eq:gSAW}) we can deduce the OAM selection
rule
\begin{equation}
\label{eq:OAMsel}
\left|l'\right|=2\left|l\right|,
\end{equation}
for $l,l' \neq 0.$ The factor of two in Eq.~(\ref{eq:OAMsel}) may be traced back to the fact that the
optomechanical interaction is bilinear in the optical field, but only linear in the acoustic field. The
selection rule of Eq.~(\ref{eq:OAMsel}) provides a path for targeting specific mechanical modes. In the
original proposal this selectivity was exploited for storing photonic OAM in mechanical vibrational
modes \cite{ShiJPhysB2013}. However, there is no analogous rigorous selection rule for $p$ and $p'$,
only simply varying degrees of overlap between modes with various radial numbers, described by the factor
$\xi_{lpp'}$ . We note that for this optomechanical platform, angular momentum is carried and exchanged in a parametric
way, and is not a dynamical variable with its own time evolution.

\section{Cavityless optomechanics}
\label{sec:CavLess}
In this section we briefly consider the interaction of mechanical degrees of freedom with electromagnetic
modes which are not confined in a resonator, but propagate freely in space. Such configurations constitute
a rapidly growing class of optomechanical systems \cite{YinIJMP2013,NeukirchCP2015}. Below we will describe
linear vibrational motion and then move on to torsional and then finally rotational motion.
\subsection{Vibrational motion}
\label{subsec:VCLO}
\begin{figure}
\begin{center}
\includegraphics[width=0.4\textwidth]{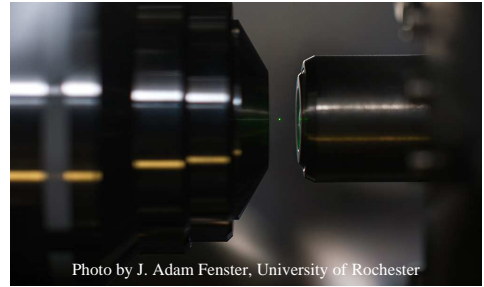}
\caption{(Color online) The system described in Section~\ref{subsec:VCLO}. An optically levitated nanoparticle
scattering light from the 1064nm trapping laser. Reproduced with permission from Prof. A. N. Vamivakas.}
\label{fig:P12}
\end{center}
\end{figure}

A typical configuration consists of a harmonically oscillating subwavelength dielectric particle trapped
optically at the focus of a freely propagating Gaussian optical beam, see Fig.~\ref{fig:P12}.
When this configuration is placed inside a liquid medium, it is usually referred to as an `optical tweezer'
\cite{Moffitt2008}. In the present case we will consider optical trapping in vacuum only. Photons scattered
by the trapped particle are used as a probe to detect its motion. At low background pressures, the particle
motion is ballistic and simple harmonic to a very good approximation. The frequency of oscillation along each
spatial direction is set by the intensity of the trapping beam.

An example of an optomechanical effect in this system is cooling of the oscillator motion, using a technique
different from that described in Section~\ref{subsec:OMCAV}. Cooling can be achieved by making the trapping
beam intensity (and therefore the mechanical trap frequency) higher when the particle is moving away from the
center of the trap, and lower when it is moving towards. Using such a technique, cooling has been implemented
from $300$K (room temperature) to $50$mK \cite{GieselerPRL2012}. The system can also be used as a thermometer
\cite{MillenNatureNano2014}, as a force sensor \cite{MoorePRL2014}, and can carry other spin-like degrees of
freedom \cite{NeukirchNP2015}. Theoretical proposals exist for creating macroscopic superpositions in such
systems \cite{ScalaPRL2013,YinPRA2013}.

\subsection{Torsional motion}
\label{subsec:TCLO}
Exchange of angular momentum between an optical field and a torsional mechanical oscillator in free space
was first seen in a classic experiment by Beth \cite{BethPR1936}. By detecting the rotation of a quarter-wave plate
suspended as a torsional oscillator due to the change in handedness of polarized light passing through it, Beth was
in fact able to experimentally show for the first time that light possesses angular momentum. The experiment
has been repeated in modern times in the optical \cite{MoothooAJP2001} as well as microwave \cite{EmilePRL2014}
domains.

\subsection{Rotational motion}
\label{subsec:RCLO}
In this section we describe several examples of the coupling of light carrying angular momentum to free
mechanical rotation.

\subsubsection{Spinning graphene}
Charged flakes of graphene have been confined in an electrostatic trap under vacuum \cite{KanePRB2010}.
The linear motion is to a good approximation harmonic along all three spatial directions. In addition, it was
found that when the flake was irradiated with circularly polarized light, it spun at about $10$ MHz. The
mechanism for angular momentum transfer was light absorption.

\subsubsection{Optomechanical gyroscope}
\label{subsubsec:Gyro}
\begin{figure}
\begin{center}
\includegraphics[width=0.5\textwidth]{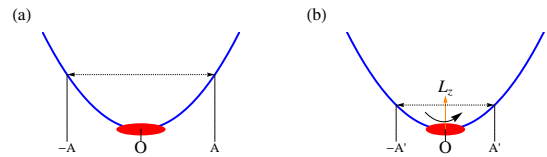}
\caption{(Color online) The system described in Section~\ref{subsubsec:Gyro}. (a) shows a trapped asymmetric
particle executing simple harmonic motion with amplitude $A$ (b) shows the same particle spinning at a constant
angular momentum $L_{z}$ about an axis transverse to the vibrational motion. The conservation of angular momentum
results in a reduced amplitude of motion $A'< A$, which may be thought of as gyroscopic cooling of vibration
\cite{YoshiNatComm2013}.}
\label{fig:P14}
\end{center}
\end{figure}

Optically trapped particles can exhibit interesting interplay between vibrational and rotational degrees
of freedom, as demonstrated using an optomechanical gyroscope \cite{YoshiNatComm2013}. The trapped nanoparticle
in this case was slightly anisotropic in shape, and the linear degrees of freedom were harmonic, as in the
examples above. As in the case of graphene, a circularly polarized light beam set the particle in rotation.
An interesting effect was then demonstrated that involved both vibration and rotation, as shown in Fig.~\ref{fig:P14}.
In the absence of rotation, the particle oscillated with a certain amplitude along each axis, as shown in
Fig.~\ref{fig:P14} (a). In the presence of rotation, however, the particle became resistant to changing the
direction of its angular momentum, as for any spinning top. This led to a decrease in the oscillation amplitude
of the vibrational particle motion,as shown in Fig.~\ref{fig:P14}(b). This effect corresponds to `gyroscopic cooling'
of vibrational motion, distinct from resonator-based (Section ~\ref{subsec:OMCAV}) and feedback (Section ~\ref{subsec:VCLO})
cooling. In the experiment cooling of other rotational modes was also observed.

\subsubsection{Rotating atoms, molecules and electrons}

For completeness we mention that there is a sizable literature associated with the transfer of optical angular
momentum to the center-of-mass motion of hot \cite{Guo2008}, cold
\cite{Tabosa1999,Muthukrishnan2002,Inoue2006,Guo2006,Pugatch2007,Moretti2009,Inoue2009,VeissierOL2013}
and degenerate
\cite{AndersenPRL2006,RyuPRL2007,WrightPRA2008,Marzlin1997,Dum1998,Nandi2004,Dutton2004,Kapale2005,Kanamoto2007,
Wright2009, Leslie2009,Paternostro2010,BeattiePRL2013} atoms, molecules \cite{Korobenko2014}, and electrons 
\cite{HandaliOE2015}. There are many phenomena of interest in this category including vortices, persistent currents, 
spin textures, and optical centrifuges. We will not describe these works in detail here, since they go back several 
years, if not decades.

\section{Conclusion and outlook}
\label{sec:Con}
In this tutorial, we have presented some introductory examples of the rich optomechanics arising from the
exchange of angular momentum between light and matter. We have attempted to relate the relevant theoretical
tools to the earlier work on vibrational optomechanics, while trying to present a useful variety of experimental
platforms on which the theory could be realized. We therefore hope our exposition will be of relevance to theorists as
well as experimentalists.

\section{Acknowledgements}
\label{sec:Ack}
We are grateful to B. Zwickl, M. Vengalattore, S. Agarwal and B. Rodenburg for useful discussions.
We would like to thank the Research Corporation for Science Advancement (Award No. 20966) and the National Science
Foundation (Award No. 1454931) for Support.

\end{document}